# Epitaxial Strain Activates Altermagnetic Spin-Splitting Torques in $RuO_2$(100)

Qi Jia[1], Seung Gyo Jeong[2], Seungjun Lee[1], Denis Tonini[1], Anand Santhosh[2], Yifei Yang[1], Xiangrui Li[1], Brahmdutta Dixit[1], Shuang Liang[2], Yu-Chia Chen[1], Tony Low[1], Bharat Jalan[2], Jian-Ping Wang[1,2,†]

[1]Department of Electrical and Computer Engineering, University of Minnesota, 200 Union St. SE, Minneapolis, MN, 55455, USA

[2]Department of Chemical Engineering and Materials Science, University of Minnesota, 421 Washington Ave. SE, Minneapolis, MN, 55455, USA

## Abstract

The altermagnetic nature of rutile $RuO_2$ remains under active debate: bulk measurements indicate a nearly nonmagnetic ground state, whereas thin-film studies have reported symmetry-dependent transport signatures consistent with altermagnetism. Here, we provide experimental evidence that altermagnetic spin splitting in $RuO_2$ is a strain-stabilized emergent state rather than an intrinsic bulk property. Angular-resolved spin–torque measurements reveal a symmetry-selected spin Hall response characteristic of altermagnetic spin splitting, which is strongest in the strained regime but progressively suppressed as the lattice relaxes toward the bulk limit. Complementary magnetic measurements further reveal enhanced coercivity and exchange-bias behavior exclusively in strained films, indicating the emergence of a strain-stabilized magnetic state. First-principles calculations reproduce the strain-dependent evolution of the Néel order and spin-split electronic structure, supporting the experimental observations. Together, these results establish altermagnetic spin splitting in $RuO_2$ as a strain-stabilized emergent state and provide a unified explanation for the long-standing discrepancy between bulk and thin-film observations.

## Introduction

Altermagnetism has emerged as a distinct magnetic phase that hosts momentum-dependent spin splitting without net magnetization, enabling spin-polarized transport in fully compensated systems[1–7]. Rutile $RuO_2$, a metallic oxide predicted to host symmetry-dependent spin-split electronic states, has become a prominent candidate for realizing such behavior at room temperature[8–10]. However, whether $RuO_2$ intrinsically hosts altermagnetic spin splitting remains under active debate.

Measurements that directly probe long-range magnetic order in bulk $RuO_2$, including neutron scattering, μSR[11,12], Mössbauer spectroscopy[13], and torque magnetometry[14], consistently indicate a nearly nonmagnetic ground state with no evidence of robust long-range order[15]. Although early neutron studies suggested small-moment antiferromagnetism[8,10], recent re-examinations indicate that these observations can instead arise from multiple scattering and structural anisotropy effects rather than intrinsic magnetic order[11–14]. Angle-resolved photoemission spectroscopy likewise fails to resolve clear spin-split electronic structures[16–18]. Together, these results point to an essentially nonmagnetic bulk ground state. In contrast, thin-film and heterostructure studies have revealed pronounced symmetry-dependent transport responses inconsistent with a fully nonmagnetic ground state. Reports of anisotropic magnetoresistance, anomalous Hall effects, magnetic tunnel junction responses, and anisotropic spin orbit torques consistently exhibit pronounced symmetry-dependent signals[19–37]. Similar symmetry-dependent features have also been reported in spectroscopic measurements[27,38–41]. These observations are broadly consistent with the theoretically predicted altermagnetic spin splitting in $RuO_2$[1,9].

This apparent discrepancy between bulk-sensitive probes and thin-film observations raises a central question: whether the reported altermagnetic spin splitting reflects an intrinsic bulk property or emerges only under specific structural conditions. Recent theoretical studies suggest that the electronic ground state and magnetic order in $RuO_2$ are highly sensitive to lattice distortion, raising the possibility that altermagnetic spin splitting may emerge only under epitaxial strain[42–45]. In particular, $RuO_2$(100) grown epitaxially on $TiO_2$(100) is predicted to provide a unique platform in which epitaxial strain stabilizes altermagnetic spin splitting, in contrast to bulk and bulk-like $RuO_2$[45]. However, in most previous studies, strain is not independently controlled but instead varies simultaneously with substrate choice, film thickness, and growth conditions, making it difficult to isolate the role of strain in stabilizing the spin-split state[46–49].

Here, we grow epitaxial $RuO_2$(100) thin films on $TiO_2$ substrates by hybrid molecular beam epitaxy (MBE), in which the anisotropic strain is systematically tuned via film thickness. On this platform, we perform symmetry-resolved ST-FMR measurements together with magnetic hysteresis characterization, allowing us to track both spin-splitting signatures and magnetic interactions across the strain-relaxation process. We find that the torque component consistent with momentum-dependent spin splitting is most pronounced in the strained regime and is progressively suppressed as the lattice relaxes toward the bulk limit. Complementary magnetic measurements further reveal strain-dependent interfacial interactions, providing independent evidence for a strain-stabilized magnetic state. Together, these results establish a thickness-controlled route for continuously tuning anisotropic epitaxial strain in $RuO_2$(100). This well-defined strain platform enables a direct examination of how magnetic order and symmetry-dependent spin transport evolve as the lattice progressively relaxes toward the bulk limit.

**Thickness-Controlled Anisotropic Strain Evolution in Epitaxial $RuO_2$**

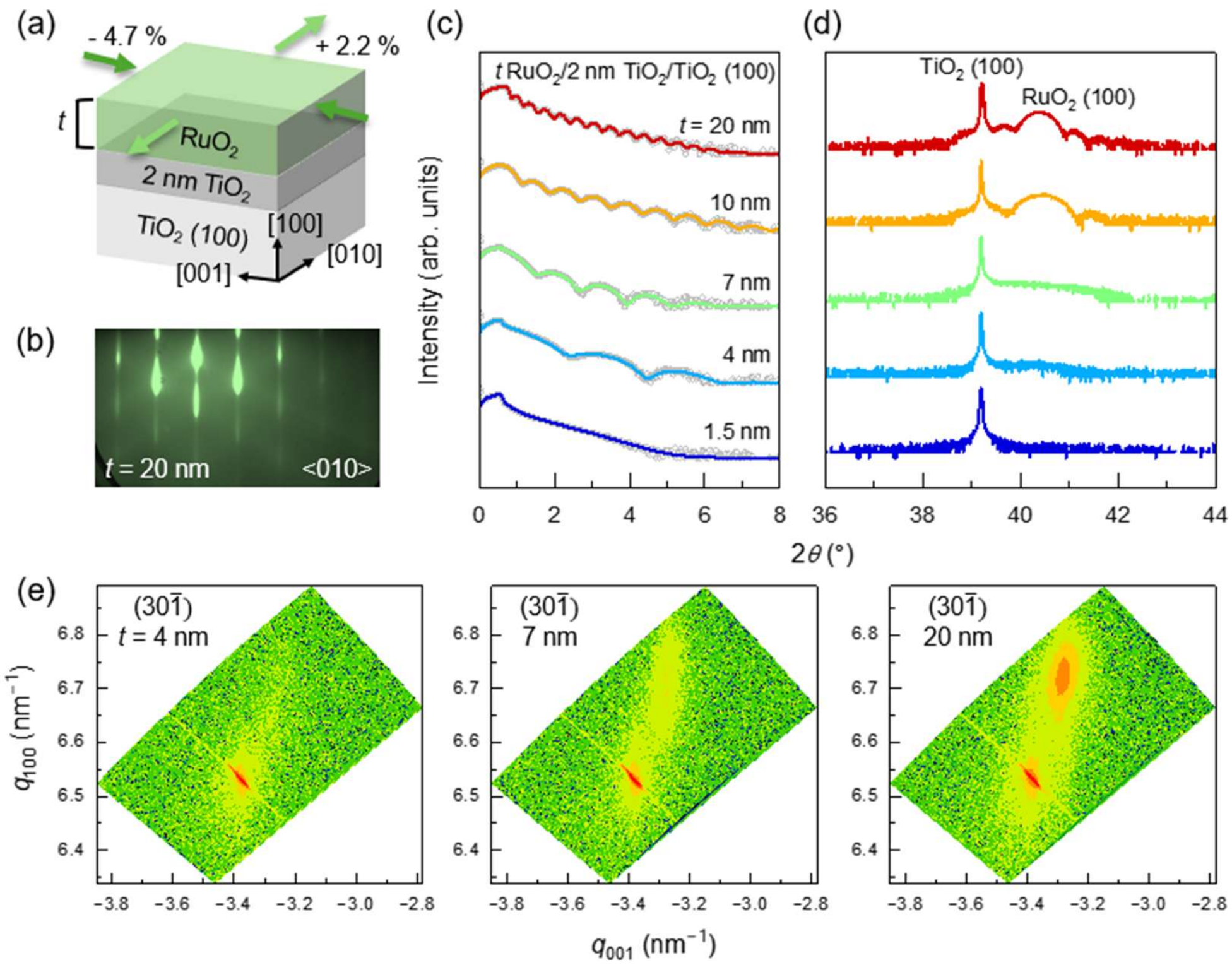

**Figure 1 | Thickness-dependent anisotropic strain evolution in epitaxial $RuO_2$(100)/$TiO_2$(100) heterostructures.** (a) Schematic illustration of epitaxial $RuO_2$(100) thin films grown on $TiO_2$(100) substrates with a 2 nm $TiO_2$ buffer layer. (b) Representative in situ reflection high-energy electron diffraction (RHEED) pattern of a 20 nm-thick $RuO_2$ film. (c) X-ray reflectivity (XRR) scans for $RuO_2$ films with thicknesses ranging from 1.5 to 20 nm. (d) X-ray diffraction (XRD) θ–2θ scans of $RuO_2$(100) films on $TiO_2$(100) from 1.5 to 20 nm. (e) Reciprocal space maps measured around the $(30\bar{1})$ reflection for 4, 7, and 20 nm $RuO_2$ films.

To establish a platform for systematically tuning epitaxial strain in $RuO_2$, we grew $RuO_2$(100) thin films on $TiO_2$(100) substrates with a 2 nm $TiO_2$ buffer layer using hybrid molecular beam epitaxy (MBE). The $TiO_2$ buffer layer was introduced to establish a uniform $RuO_2$/$TiO_2$ interface independent of substrate-to-substrate surface variations. For the rutile (100) orientation, the lattice mismatches between $RuO_2$ and $TiO_2$ are highly anisotropic, reaching −4.7% along the [001] direction and +2.2% along [010], as illustrated in Fig. 1a. This anisotropic lattice mismatch enables distinct relaxation behavior along the two in-plane crystallographic directions as the film thickness increases[50].

The epitaxial growth and interface quality of the $RuO_2$ films were first confirmed by in situ reflection high-energy electron diffraction (RHEED) and X-ray reflectivity (XRR) measurements. Representative RHEED patterns for a 20 nm-thick $RuO_2$ film are shown in Fig. 1b, exhibiting streaky diffraction features characteristic of high-quality epitaxial growth. XRR scans in Fig. 1c display pronounced Kiessig fringes persisting up to ~8°, indicative of atomically smooth interfaces and excellent thickness uniformity throughout the film series. Quantitative fitting of the XRR spectra (solid lines) further confirms precise thickness control of the $RuO_2$ films over the thickness range from 1.5 to 20 nm. X-ray diffraction (XRD) θ–2θ scans shown in Fig. 1d reveal pronounced rutile (100) Bragg reflections from both the $TiO_2$ substrate and $RuO_2$ films, accompanied by clear Laue oscillations, indicating high crystalline coherence throughout the thickness series.

To directly probe the strain evolution, reciprocal space maps (RSMs) measured around the $(30\bar{1})$ reflection are presented in Fig. 1e for $RuO_2$ films with thicknesses of 4, 7, and 20 nm. While films below ~ 4 nm remain close to the coherently strained limit, a second diffraction feature emerges near $q_{001} \sim -3.29$ nm$^{-1}$ once the film thickness reaches ~ 7 nm. The appearance of this additional peak indicates the coexistence of coherently strained and partially relaxed lattice regions within the film. As the thickness further increases, the intensity of the relaxed component grows progressively, revealing a continuous release of epitaxial strain along the [001] direction. In contrast, reciprocal space maps measured around the (310) reflection (Fig. S1) show that the $RuO_2$ films remain coherently locked to the $TiO_2$ substrate along the [010] direction even at 20 nm thickness, establishing a highly anisotropic strain relaxation process in $RuO_2$(100)/$TiO_2$(100) heterostructures.

The [001] lattice constants associated with the relaxed diffraction component were estimated to be 3.047 Å and 3.054 Å for the 7 nm and 20 nm $RuO_2$ films, respectively, with an uncertainty of ±0.005 Å. The extracted lattice constants progressively approach the relaxed bulk value with increasing thickness, further confirming the gradual reduction of epitaxial strain along [001]. Together, these results establish a thickness-controlled route for continuously tuning anisotropic epitaxial strain in $RuO_2$(100), providing a well-defined platform for examining its impact on the magnetic and spin-transport properties discussed below.

**Symmetry-Selective Spin Transport in Strained $RuO_2$**

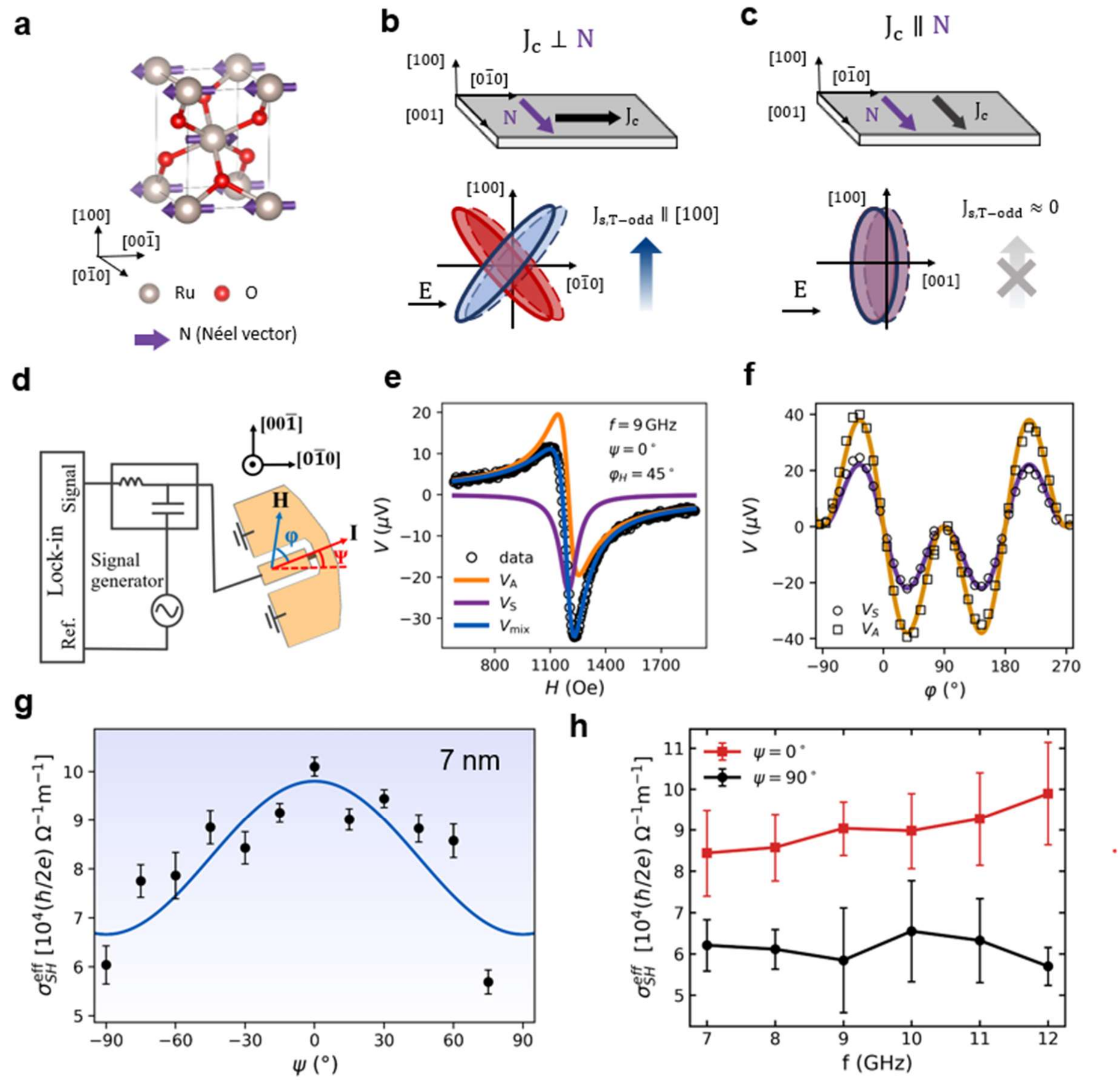

**Figure 2. Symmetry-dependent T-odd spin Hall conductivity in epitaxial $RuO_2$.** (a) Crystal structure of rutile $RuO_2$ showing the Néel vector (N). (b,c) Symmetry analysis of the T-odd spin Hall effect for ($J_c \perp$ N) and ($J_c \parallel$ N), respectively. (d) Device geometry and measurement configuration for spin-torque ferromagnetic resonance (ST-FMR), defining the current direction ψ relative to the crystal axes and the magnetic field angle φ. (e) Representative mixing-voltage spectrum $V_{\mathrm{mix}}$ as a function of magnetic field for ψ = 0° and φ = 45°, decomposed into symmetric ($V_S$) and antisymmetric ($V_A$) components. (f) Angular dependence of the symmetric and antisymmetric components as a function of φ. (g) Effective Spin Hall conductivity (SHC) as a function of current direction ψ, exhibiting a pronounced $\cos 2\psi$ dependence with maximum at ψ = 0°, consistent with T-odd spin splitting contribution. (h) Frequency dependence of the extracted effective SHC for ψ = 0° and 90°, demonstrating that the angular anisotropy is robust and independent of excitation frequency.

Fig. 2a shows the rutile crystal structure of $RuO_2$, where the Néel vector is aligned along the [001] direction. Under this symmetry, altermagnetic spin splitting is expected to generate an additional spin-polarization component when the charge current is applied along the [010] direction, i.e., perpendicular to the Néel vector (Jc ⊥ N), as illustrated in Fig. 2b. In contrast, this spin-splitting contribution is symmetry-forbidden when the current is applied along the [001] direction, parallel to the Néel vector (Jc ∥ N), leaving only the conventional response (Fig. 2c). We refer to this symmetry-selective spin-polarization contribution as the T-odd component, whereas the conventional contribution is denoted as the T-even component. The appearance of a finite T-odd response therefore provides a measurable transport signature associated with the symmetry-selective spin-splitting contribution.

To test this symmetry prediction experimentally, we first focus on a representative strained $RuO_2$(100) film with a thickness of 7 nm and perform spin-torque ferromagnetic resonance (ST-FMR) measurements on patterned $RuO_2$/Py bilayer devices. The measurement geometry is illustrated in Fig. 2d, where the current direction $\psi$ is defined relative to the [010] crystallographic direction and systematically varied by rotating the device orientation in 15° increments, thereby allowing the symmetry dependence of the generated spin current to be directly probed. The external magnetic field is applied at an in-plane angle $\varphi$ relative to the current direction, allowing the torque symmetry to be independently evaluated for each current direction $\psi$. A representative mixing-voltage spectrum measured on a 7 nm $RuO_2$ sample at $\psi = 0°$ and $\varphi = 45°$ is shown in Fig. 2e. The signal is decomposed into symmetric and antisymmetric Lorentzian components using a standard ST-FMR fitting procedure (see Methods). Fig. 2f shows the angular dependence of the extracted symmetric and antisymmetric voltage components for the $\psi = 0°$ device. The experimental data are well reproduced by the standard ST-FMR model using the leading $cos\phi \sin 2\phi$ term. The excellent agreement confirms that the measured response follows the expected ST-FMR angular symmetry, supporting the extraction procedure described in Methods.

We next examine the dependence of the extracted effective SHC on the current direction $\psi$. As shown in Fig. 2g, the effective SHC exhibits a pronounced twofold angular dependence that is well described by a $\cos 2\psi$ symmetry. The SHC reaches its maximum for the current applied along [010] (Jc ⊥ N) and decreases toward [001] (Jc ∥ N). This behavior is consistent with the symmetry-selective contribution illustrated in Fig. 2b,c, where the T-odd spin-splitting contribution is allowed for Jc ⊥ N but forbidden for Jc ∥ N. To assess the reproducibility of the observed anisotropy, we further examine the extracted SHC over the frequency range from 7 to 12 GHz. As shown in Fig. 2h, the effective SHC remains consistently larger for $\psi = 0°$ than for $\psi = 90°$ across the entire frequency range, indicating that the symmetry-selective response persists over the measured frequency window.

## Thickness dependence of spin splitting torque

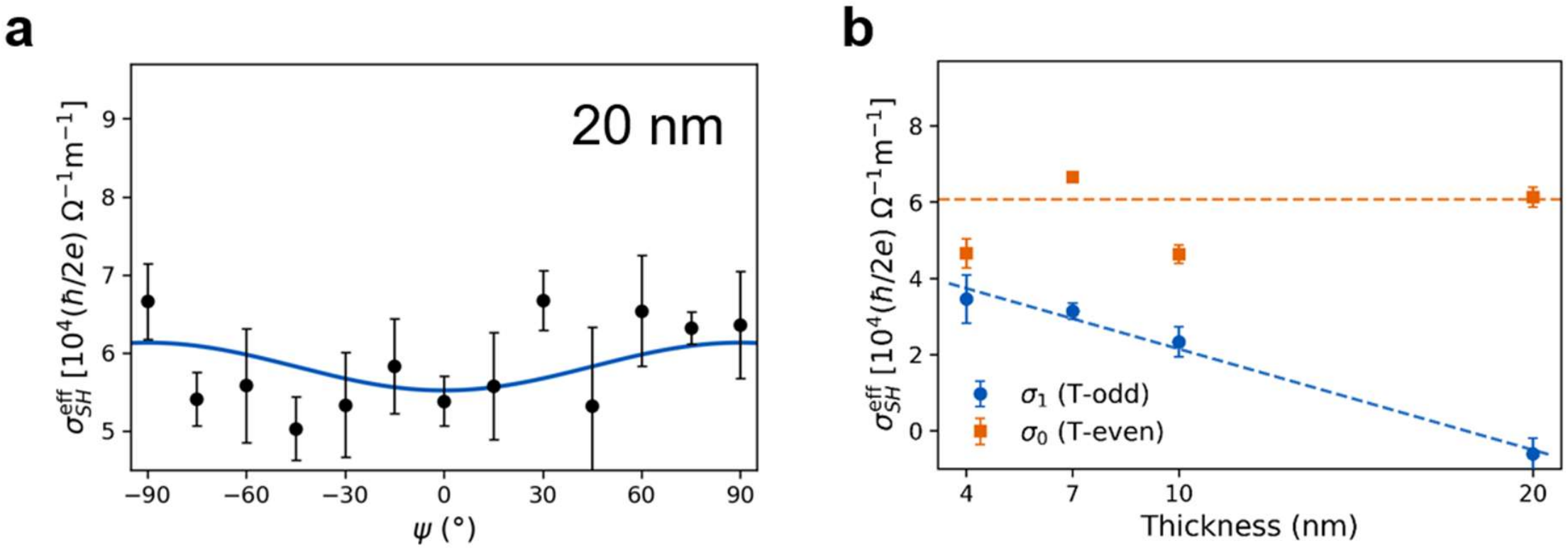

**Figure 3 | Thickness dependence of symmetry-selective spin response in $RuO_2$(100).** (a) Effective Spin Hall conductivity (SHC) as a function of current direction ψ for $RuO_2$ films with thicknesses of 20 nm. (b) Thickness dependence of the extracted SHC components. The constant term remains nearly unchanged across the thickness range, while the amplitude of the angular (T-odd) component decreases continuously and approaches zero in the thick-film limit.

Having established the presence of a symmetry-selective spin response in the strained regime, we next investigate its evolution with film thickness. Representative angular-dependent measurements for a 20 nm $RuO_2$ film are shown in Fig. 3a. In contrast to the pronounced angular dependence observed in the strained 7 nm film (Fig. 2g), the effective SHC of the 20 nm film exhibits only a weak angular dependence on the current direction ψ. The disappearance of the characteristic positive cos2ψ anisotropy indicates that the symmetry-selective spin response is strongly suppressed as the film approaches the relaxed limit. The same behavior is observed across different excitation frequencies for the 20 nm sample (Fig. S5), confirming that the suppression of the T-odd contribution is robust against measurement conditions.

To quantify this evolution, the angular dependence of the effective SHC for each thickness is fitted using Acos2ψ + C, where (C) represents the angle-independent background contribution (T-even component), and (2A), corresponding to the peak-to-valley difference of the angular modulation, is taken as the magnitude of the T-odd component. The extracted T-even and T-odd contributions are summarized in Fig. 3b. While the T-even component remains approximately unchanged across the thickness series, the T-odd component decreases systematically with increasing thickness and approaches zero in the thick-film limit.

Because the T-odd term originates from the symmetry-selective spin-splitting contribution introduced in Fig. 2, its suppression indicates a progressive collapse of the spin-splitting-related response, while the conventional spin-current response remains largely unchanged. The disappearance of the T-odd contribution in thick films indicates that the spin-splitting-related symmetry-selective response is not an intrinsic property of relaxed $RuO_2$, but instead emerges only within the strained epitaxial regime. The systematic suppression of the T-odd component closely follows the thickness-dependent strain relaxation established in Fig. 1, establishing a strong correlation between epitaxial strain and the symmetry-selective spin response.

## FORC characterization

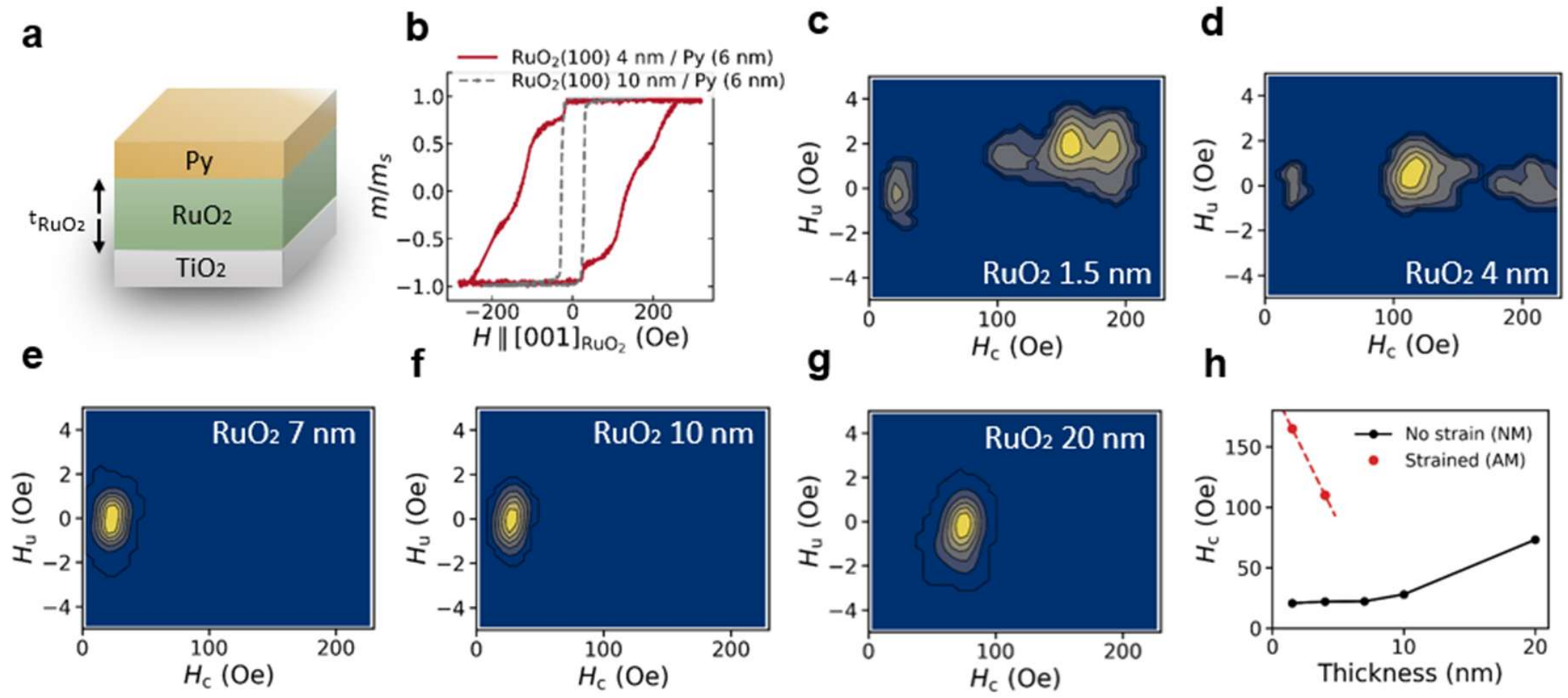

**Figure 4 | Thickness-dependent evolution of interfacial magnetic state revealed by FORC analysis.** (a) Schematic illustration of the $RuO_2$/Py heterostructure grown on $TiO_2$(100), where the $RuO_2$ thickness is varied to tune the epitaxial strain. (b) Representative M–H loops measured at 10 K with the magnetic field applied along the [001] direction for $RuO_2$/Py bilayers, comparing thin (4 nm) and thicker (10 nm) $RuO_2$ layers. (c–g) First-order reversal curve (FORC) distributions for $RuO_2$ thicknesses of 1.5 nm (c), 4 nm (d), 7 nm (e), 10 nm (f), and 20 nm (g). (h) Thickness dependence of the extracted coercive fields. The low-coercivity component evolves smoothly with thickness and represents a baseline response dominated by the Py layer. In contrast, the high-coercivity component appears only in the strained regime and disappears upon strain relaxation, identifying it as a strain-stabilized interfacial magnetic contribution.

To further probe the magnetic origin of the strain-dependent transport response, we examine the magnetization behavior of $RuO_2$/Py bilayers. The M–H loops measured at room temperature reveal a clear in-plane magnetic anisotropy, with the easy axis aligned along the [001] direction of $RuO_2$ (see Fig. S6). This anisotropy is consistent across all thicknesses. While it is consistent with a preferred alignment of an underlying magnetic order along the [001] direction, it does not by itself distinguish the microscopic origin, as it may also arise from interfacial anisotropy[51].

Upon cooling to 10 K under zero-field conditions, a marked thickness-dependent contrast in the magnetization behavior is observed. Representative M–H loops along [001] are shown in Fig. 4a. The Py layer on thicker $RuO_2$ (10 nm) exhibits sharp, nearly single-step switching with relatively small coercivity, whereas that on thinner $RuO_2$ (4 nm) shows substantially enhanced coercivity accompanied by distinct multi-step switching behavior. Because all samples contain identical Py layers deposited under identical conditions, the observed differences originate from the underlying $RuO_2$ layer.

To resolve this evolution, we perform first-order reversal curve (FORC) measurements as a function of $RuO_2$ thickness (see Methods). The FORC distribution maps switching events into a two-dimensional space of coercive field ($H_c$) and interaction field ($H_u$), where distinct peaks correspond to independent switching populations, with $H_c$ characterizing the switching barrier and $H_u$ capturing the local interaction field. The results are shown in Fig. 4b–f. A low-coercivity population persists across all thicknesses and remains centered at $H_u \approx 0$. In contrast, a distinct high-coercivity population emerges exclusively in the strained regime and disappears upon increasing thickness beyond 7 nm. In the 1.5 nm film, this population exhibits a pronounced shift along $H_u$, revealing a finite exchange bias field.

Most importantly, the coexistence of a finite exchange bias and an enhanced high-coercivity switching population in the strained regime demonstrates that epitaxial strain stabilizes an additional interfacial magnetic state that is absent in relaxed $RuO_2$[52]. Upon strain relaxation, this magnetic state disappears over a narrow thickness range between 4 and 7 nm, revealing a clear magnetic crossover between strained and relaxed $RuO_2$. In contrast, the low-coercivity switching population persists throughout the entire thickness series and evolves continuously with thickness, indicating that it represents the baseline magnetic response of the $RuO_2$/Py bilayer (Fig. 4g). Notably, the emergence of this strain-stabilized magnetic state coincides with the appearance of the T-odd transport response identified in Fig. 3, suggesting a common strain-dependent origin. Although the high-coercivity population disappears over a narrower thickness range than the transport signal, the two probes are expected to sample different length scales, with FORC primarily sensitive to interfacial magnetic interactions and transport measurements integrating contributions from a larger volume of the $RuO_2$ layer.

Temperature-dependent coercivity measurements (Fig. S7) further distinguish the high-Hc switching population from the baseline low-Hc response. The anomalous high-Hc component exhibits a much stronger temperature dependence yet remains observable up to room temperature, demonstrating that the additional magnetic state persists under the same conditions where the T-odd transport response is measured.

**Theoretical calculation**

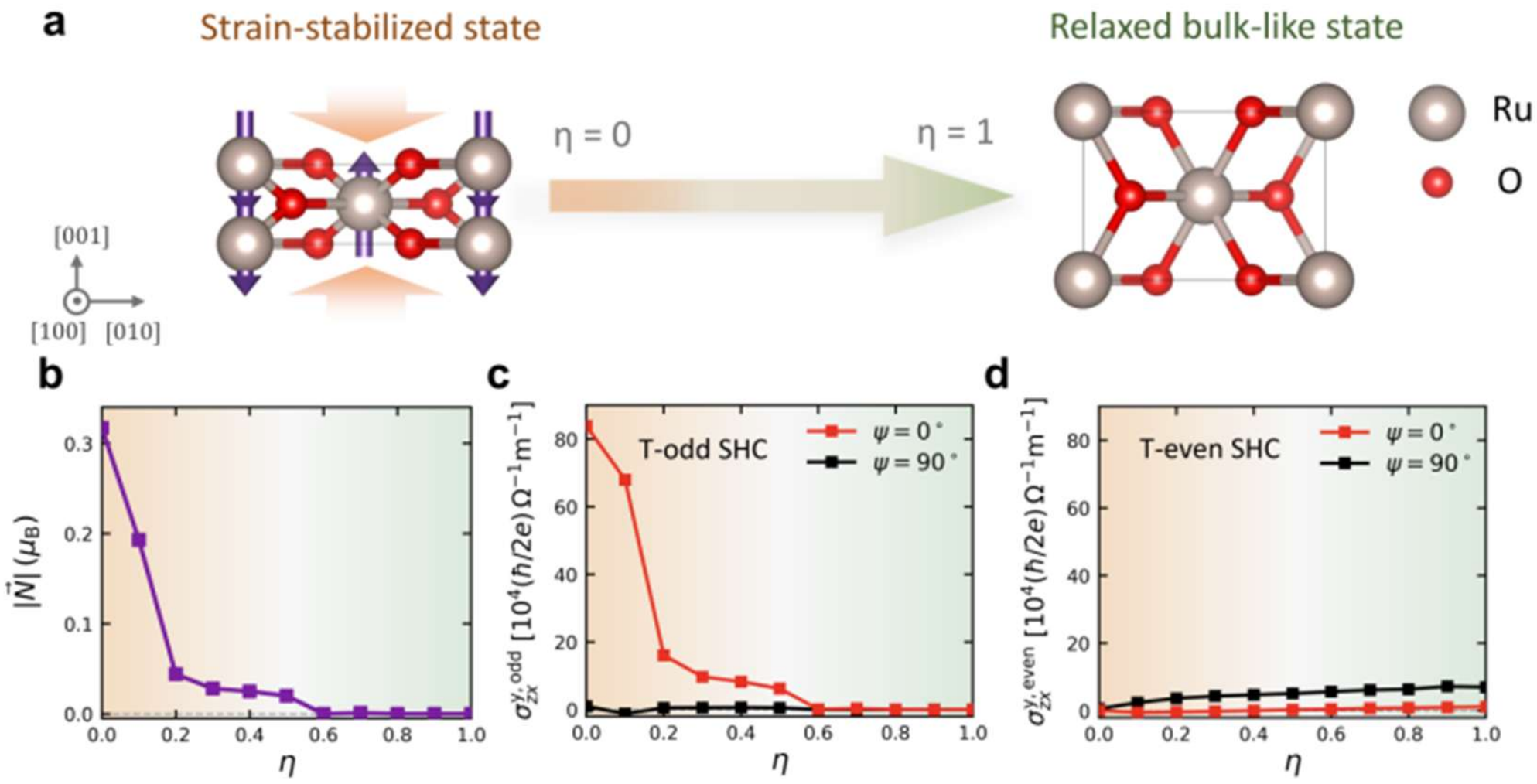

**Figure 5 | Strain-dependent magnetic order and spin Hall conductivity of the (100) $RuO_2$ thin film from first-principles calculations.** (a) Schematic illustration of the strain-driven crossover from a strain-stabilized magnetic state to a relaxed bulk-like state, parameterized by the strain relaxation factor η. The strained state (η = 0) hosts a finite magnetic order, whereas the magnetic order is progressively suppressed

as η approaches unity. (b) Magnitude of the Néel vector $|\vec{N}| = |\vec{\mu}_{\mathrm{Ru}_1} - \vec{\mu}_{\mathrm{Ru}_2}|/2$ as a function of the strain parameter $\eta$, where $\eta = 0$ corresponds to the coherently strained limit matching the $TiO_2$ substrate and $\eta = 1$ to the fully relaxed $RuO_2$ lattice along [001]. (c) T-odd and (d) T-even of components the spin Hall conductivity $(\sigma_{zx}^{y,\mathrm{odd}}$ and $\sigma_{zx}^{y,\mathrm{even}})$ as a function of ε for current directions ψ = 90° (black) and ψ = 0° (red), where ψ indicates a rotation angle of $x$ from the [010] crystallographic axis.

To clearly understand the microscopic origin of the thickness-dependent transport properties, we performed first-principles calculations based on density functional theory (DFT). We first investigated the Néel vector magnitude, defined as $|\vec{N}| = |\vec{\mu}_{\mathrm{Ru}_1} - \vec{\mu}_{\mathrm{Ru}_2}|/2$, where $\vec{\mu}_{\mathrm{Ru}_1}$ and $\vec{\mu}_{\mathrm{Ru}_2}$ are two opposite local magnetic moments of the Ru atoms. The strain dependence was introduced through $a^{001} = (1-\eta)a^{001}_{\mathrm{TiO}_2} + \eta a^{001}_{\mathrm{Ru}\ 2}$, where $\eta$ denotes the strain parameter. In our calculations, the other in-plane lattice constant along the [010] direction was fixed to the $TiO_2$ substrate value, whereas the out-of-plane lattice constant along [100] was fully relaxed, consistent with experimental observations.

Fig. 5a shows the calculated $|\vec{N}|$ as a function of $\eta$. The Néel vector remains essentially zero for $\eta > 0.6$, but increases rapidly at larger strain, reaching ~ 0.3 $\mu_\mathrm{B}$ per Ru site near the fully-strained limit. This indicates that the relaxed lattice supports a nonmagnetic ground state, whereas magnetic order in (100)-oriented $RuO_2$ is strain-activated. Fig. 5b and 5c show the strain dependence of the representative T-odd ($\sigma_{zx}^{y,\mathrm{odd}}$) and T-even ($\sigma_{zx}^{y,\mathrm{even}}$) components of the SHC tensor obtained from linear-response calculations. (see Methods) The T-odd component closely tracks the onset of $|\vec{N}|$ for ψ = 0°, becoming sizable only for $\eta \leq 0.6$. By contrast, it remains strongly suppressed for ψ = 90° due to the symmetry of the altermagnetic spin-split bands. The small residual T-odd response for ψ = 90° near $\eta = 0$ originates from the spin-orbit coupling. In contrast, the T-even component is finite across the entire strain range and varies smoothly with $\eta$, reflecting its spin-orbit-driven origin, which is less sensitive to magnetic order. These two contributions, therefore, have distinct microscopic origins and strain dependences.

This separation provides a natural framework for the thickness-dependent angular response in ST-FMR. The constant term in the A cos2ψ + const. fits (Fig. 3a), which is nearly thickness-independent and survives in the relaxed 20 nm film, is consistent with the T-even contribution, whereas the angular component A, which decays continuously with thickness, tracks the strain-induced T-odd contribution.

We note that the in-plane anisotropy of the rutile (100) surface does not strictly forbid a ψ-dependent T-even SHC, and this contribution may itself vary with strain and Fermi level. Nevertheless, two key observations identify the T-odd channel as the dominant origin of the observed angular contrast. First, the angular component extracted from the ST-FMR measurements vanishes in the relaxed limit, coinciding with the collapse of the Néel order and the disappearance of the calculated T-odd SHC. Second, this disappearance occurs simultaneously with the loss of the additional magnetic state identified by the FORC measurements (Fig. 4), providing an independent magnetic signature of the strain-stabilized phase. Although a strain-dependent T-even anisotropy may also contribute to the angular response, it cannot naturally explain the simultaneous disappearance of the magnetic signatures and the angular transport response upon strain relaxation. Taken together, these results provide compelling evidence that epitaxial strain stabilizes the altermagnetic spin-split state in $RuO_2$(100), giving rise to the symmetry-selective spin transport observed experimentally.

**Discussion**

The present results establish epitaxial strain as the key parameter governing the emergence of the altermagnetic spin-split state in $RuO_2$. As the lattice progressively relaxes toward the bulk limit, the T-odd transport response, the strain-stabilized magnetic state, and the calculated Néel order are all systematically suppressed. This unified evolution provides a direct explanation for the long-standing discrepancy between bulk-sensitive measurements and thin-film observations in $RuO_2$, indicating that the experimentally observed altermagnetic behavior is a strain-stabilized phenomenon of epitaxial thin films rather than a universal property of the relaxed bulk crystal. More broadly, our findings demonstrate that epitaxial strain can qualitatively reshape competing magnetic and electronic ground states, providing an effective route for engineering symmetry-dependent spin transport phenomena in altermagnetic materials.

**References**

1. Šmejkal, L., González-Hernández, R., Jungwirth, T. & Sinova, J. Crystal time-reversal symmetry breaking and spontaneous Hall effect in collinear antiferromagnets. *Sci. Adv.* **6**, eaaz8809 (2020).
2. Song, C. *et al.* Altermagnets as a new class of functional materials. *Nat. Rev. Mater.* **10**, 473–485 (2025).
3. Tamang, R., Gurung, S., Rai, D. P., Brahimi, S. & Lounis, S. Altermagnetism and Altermagnets: A Brief Review. *Magnetism* **5**, 17 (2025).
4. Šmejkal, L., Sinova, J. & Jungwirth, T. Emerging Research Landscape of Altermagnetism. *Phys. Rev. X* **12**, 040501 (2022).
5. Šmejkal, L., Sinova, J. & Jungwirth, T. Beyond Conventional Ferromagnetism and

Antiferromagnetism: A Phase with Nonrelativistic Spin and Crystal Rotation Symmetry. *Phys. Rev. X* **12**, 031042 (2022).

6. Bai, H. *et al.* Antiferromagnetism: An efficient and controllable spin source. *Appl. Phys. Rev.* **9**, 041316 (2022).
7. Dai, J. *et al.* Research progress and future prospects of altermagnets. *Sci. Sin. Phys. Mech. Astron.* **56**, 227503 (2025).
8. Berlijn, T. *et al.* Itinerant Antiferromagnetism in ${\mathrm{RuO}}_{2}$. *Phys. Rev. Lett.* **118**, 077201 (2017).
9. Ahn, K.-H., Hariki, A., Lee, K.-W. & Kuneš, J. Antiferromagnetism in ${\mathrm{RuO}}_{2}$ as $d$-wave Pomeranchuk instability. *Phys. Rev. B* **99**, 184432 (2019).
10. Zhu, Z. H. *et al.* Anomalous Antiferromagnetism in Metallic ${\mathrm{RuO}}_{2}$ Determined by Resonant X-ray Scattering. *Phys. Rev. Lett.* **122**, 017202 (2019).
11. Keßler, P. *et al.* Absence of magnetic order in $RuO_2$: insights from μSR spectroscopy and neutron diffraction. *Npj Spintron.* **2**, 50 (2024).
12. Hiraishi, M. *et al.* Nonmagnetic Ground State in ${\mathrm{RuO}}_{2}$ Revealed by Muon Spin Rotation. *Phys. Rev. Lett.* **132**, 166702 (2024).
13. Yumnam, G. *et al.* Constraints on magnetism and correlations in $RuO_2$ from lattice dynamics and Mössbauer spectroscopy. *Cell Rep. Phys. Sci.* **6**, 102852 (2025).
14. Qian, T. *et al.* Determining the nature of magnetism in altermagnetic candidate ${\mathrm{RuO}}_{2}$. *Phys. Rev. Res.* **8**, L012072 (2026).
15. Wu, Z. *et al.* Fermi Surface of ${\mathrm{RuO}}_{2}$ Measured by Quantum Oscillations.

*Phys. Rev. X* **15**, 031044 (2025).

16. Liu, J. *et al.* Absence of Altermagnetic Spin Splitting Character in Rutile Oxide ${\mathrm{RuO}}_{2}$. *Phys. Rev. Lett.* **133**, 176401 (2024).
17. Lin, Z. *et al.* Bulk band structure of Ru O 2 measured with soft x-ray angle-resolved photoemission spectroscopy. *Phys. Rev. B* **111**, 134450 (2025).
18. Osumi, T. *et al.* Spin-degenerate bulk bands and topological surface states associated with Dirac nodal lines in ${\mathrm{RuO}}_{2}$. *Phys. Rev. B* **113**, 085116 (2026).
19. Bose, A. *et al.* Tilted spin current generated by the collinear antiferromagnet ruthenium dioxide. *Nat. Electron.* **5**, 267–274 (2022).
20. Bai, H. *et al.* Observation of Spin Splitting Torque in a Collinear Antiferromagnet ${\mathrm{RuO}}_{2}$. *Phys. Rev. Lett.* **128**, 197202 (2022).
21. Feng, Z. *et al.* An anomalous Hall effect in altermagnetic ruthenium dioxide. *Nat. Electron.* **5**, 735–743 (2022).
22. Tschirner, T. *et al.* Saturation of the anomalous Hall effect at high magnetic fields in altermagnetic RuO2. *APL Mater.* **11**, 101103 (2023).
23. Guo, Y. *et al.* Direct and Inverse Spin Splitting Effects in Altermagnetic RuO2. *Adv. Sci.* **11**, 2400967 (2024).
24. Karube, S. *et al.* Observation of Spin-Splitter Torque in Collinear Antiferromagnetic ${\mathrm{RuO}}_{2}$. *Phys. Rev. Lett.* **129**, 137201 (2022).
25. Jung, H. *et al.* Reversible Spin Splitting Effect in Altermagnetic RuO2 Thin Films. *Nano Lett.* **25**, 16985–16991 (2025).
26. Chen, H. *et al.* Spin-Splitting Magnetoresistance in Altermagnetic RuO2 Thin Films. *Adv.*

*Mater.* **37**, 2507764 (2025).

27. Zhang, Y. *et al.* Electrical manipulation of spin splitting torque in altermagnetic RuO2. *Nat. Commun.* **16**, 5646 (2025).
28. Jia, Q. *et al.* Correction of out-of-plane spin torque extracted from second harmonic Hall measurements: Role of antisymmetric planar Hall effects. *Phys. Rev. B* **112**, 174409 (2025).
29. Wu, Y. *et al.* Field-free spin–orbit switching of canted magnetization in Pt/Co/Ru/RuO2(101) multilayers. *Appl. Phys. Lett.* **126**, 012401 (2025).
30. Feng, X. *et al.* Incommensurate Spin Density Wave in Antiferromagnetic ${\mathrm{RuO}}_{2}$ Evinced by Abnormal Spin Splitting Torque. *Phys. Rev. Lett.* **132**, 086701 (2024).
31. Bai, H. *et al.* Efficient Spin-to-Charge Conversion via Altermagnetic Spin Splitting Effect in Antiferromagnet ${\mathrm{RuO}}_{2}$. *Phys. Rev. Lett.* **130**, 216701 (2023).
32. Liu, Y. *et al.* Inverse Altermagnetic Spin Splitting Effect-Induced Terahertz Emission in RuO2. *Adv. Opt. Mater.* **11**, 2300177 (2023).
33. Chu, R. Y. *et al.* Third-Order Nonlinear Hall Effect in Altermagnet ${\mathrm{RuO}}_{2}$. *Phys. Rev. Lett.* **135**, 216703 (2025).
34. Zhang, Y. *et al.* Simultaneous High Charge-Spin Conversion Efficiency and Large Spin Diffusion Length in Altermagnetic RuO2. *Adv. Funct. Mater.* **34**, 2313332 (2024).
35. Dai, J. *et al.* Thermal imaging of 180o domain contrast in altermagnetic RuO2. *Natl. Sci. Rev.* nwag203 (2026) doi:10.1093/nsr/nwag203.
36. Zhou, X. *et al.* Crystal Thermal Transport in Altermagnetic ${\mathrm{RuO}}_{2}$. *Phys. Rev. Lett.* **132**, 056701 (2024).

37. Liao, C.-T., Wang, Y.-C., Tien, Y.-C., Huang, S.-Y. & Qu, D. Separation of Inverse Altermagnetic Spin-Splitting Effect from Inverse Spin Hall Effect in ${\mathrm{RuO}}_{2}$. *Phys. Rev. Lett.* **133**, 056701 (2024).

38. He, C. *et al.* Evidence for single variant in altermagnetic RuO2(101) thin films. *Nat. Commun.* **16**, 8235 (2025).

39. Jeong, S. G. *et al.* Altermagnetic polar metallic phase in ultrathin epitaxially strained RuO2 films. *Proc. Natl. Acad. Sci.* **123**, e2526641123 (2026).

40. Zhang, Y.-C. *et al.* Probing the Néel Order in Altermagnetic RuO2 Films via X-Ray Magnetic Linear Dichroism. *Chin. Phys. Lett.* **42**, 027301 (2025).

41. Fedchenko, O. *et al.* Observation of time-reversal symmetry breaking in the band structure of altermagnetic RuO2. *Sci. Adv.* **10**, eadj4883 (2024).

42. Smolyanyuk, A., Mazin, I. I., Garcia-Gassull, L. & Valentí, R. Fragility of the magnetic order in the prototypical altermagnet ${\mathrm{RuO}}_{2}$. *Phys. Rev. B* **109**, 134424 (2024).

43. Forte, J. D. S. *et al.* Strain Engineering of Altermagnetic Symmetry in Epitaxial RuO$_2$ Films. Preprint at https://doi.org/10.48550/arXiv.2510.26581 (2026).

44. Qian, Z., Yang, Y., Liu, S. & Wu, C. Fragile unconventional magnetism in ${\mathrm{RuO}}_{2}$ by proximity to Landau-Pomeranchuk instability. *Phys. Rev. B* **111**, 174425 (2025).

45. Lee, S., Jeong, S. G., Wang, J.-P., Jalan, B. & Low, T. Strain-Driven Altermagnetic Spin-Splitting Effect in RuO2. *Nano Lett.* https://doi.org/10.1021/acs.nanolett.6c00768 (2026) doi:10.1021/acs.nanolett.6c00768.

46. Wang, Y.-C. *et al.* Absence of Transport Altermagnetic Spin-Splitting Effect in RuO2. *Nano*

*Lett.* **26**, 2548–2554 (2026).

47. Qiao, J. *et al.* Enhanced Damping-Like Torque through Strain Modulation in RuO2. *Nano Lett.* **26**, 5873–5881 (2026).

48. Yang, D. *et al.* Ferromagnetic Interface Engineering of Spin-Charge Conversion in ${\mathrm{RuO}}_{2}$. *Phys. Rev. Lett.* **136**, 026702 (2026).

49. Akashdeep, A. *et al.* Surface-localized magnetic order in RuO2 thin films revealed by low-energy muon probes. *Appl. Phys. Lett.* **128**, 022406 (2026).

50. Wadehra, N. *et al.* Strain-induced superconductivity in RuO2(100) thin-films. *Commun. Mater.* **6**, 135 (2025).

51. Wu, Y. *et al.* Perpendicular magnetic anisotropy in permalloy ultrathin film grown on RuO2(101) surface. *Appl. Phys. Lett.* **124**, 162402 (2024).

52. Nogués, J. & Schuller, I. K. Exchange bias. *J. Magn. Magn. Mater.* **192**, 203–232 (1999).

## ACKNOWLEDGMENTS

This work was supported partially by the National Science Foundation (NSF) through the University of Minnesota MRSEC under Award No. DMR-2011401. This project was also supported in part by NSF ASCENT program TUNA: No. 2230963. Parts of this work were carried out in the Characterization Facility, University of Minnesota, which receives partial support from the NSF through the MRSEC (Award No. DMR-2011401) and the NNCI (Award No. ECCS-2025124) programs. Portions of this work were conducted in the Minnesota Nano Center, which is supported by the NSF through the NNCI under Award No. ECCS-2025124. Film synthesis and structural characterizations (S.G.J. and B.J.) were supported by the U.S. Department of Energy through grant Nos. DE-SC0020211, and (partly) DE-SC0024710. Film growth was performed using instrumentation funded by AFOSR DURIP awards FA9550-18-1-0294 and FA9550-23-1-0085. S.G.J., B.J., S.L., and T.L. also benefited from the Air Force Office of Scientific Research Multi University Research Initiative (AFOSR MURI, Award No. FA9550-25-1-0262). A.S. acknowledges support from the NSF through the MRSEC program under Award No. DMR-2011401.